\documentclass[twocolumn,showpacs,aps,prd,superscriptaddress]{revtex4}

\usepackage{graphicx}
\usepackage{dcolumn}
\usepackage{amsmath}
\usepackage{epsfig}

\usepackage{times}
\usepackage{here}
\usepackage{lscape}
\usepackage{rotating}
\usepackage{wrapfig}
\usepackage{psfrag}
\usepackage{color}
\usepackage{rotate}
\usepackage{amssymb}
\usepackage{multirow}

\RequirePackage{xspace}

\usepackage{relsize}
\def\babar{\mbox{\slshape B\kern-0.1em{\smaller A}\kern-0.1em
    B\kern-0.1em{\smaller A\kern-0.2em R}}}

\def\epem       {\ensuremath{e^+e^-}\xspace}

\def\t     {\ensuremath{t}\xspace}

\def\pip   {\ensuremath{\pi^+}\xspace}

\def\Kbar  {\kern 0.2em\overline{\kern -0.2em K}{}\xspace}

\def\Kz    {\ensuremath{K^0}\xspace}
\def\Kzb   {\ensuremath{\Kbar^0}\xspace}
\def\KzKzb {\ensuremath{\Kz \kern -0.16em \Kzb}\xspace}
\def\Kp    {\ensuremath{K^+}\xspace}
\def\Km    {\ensuremath{K^-}\xspace}

\def\KpKm  {\ensuremath{\Kp \kern -0.16em \Km}\xspace}

\def\Dbar    {\kern 0.2em\overline{\kern -0.2em D}{}\xspace}

\def\Dz      {\ensuremath{D^0}\xspace}
\def\Dzb     {\ensuremath{\Dbar^0}\xspace}
\def\DzDzb   {\ensuremath{\Dz {\kern -0.16em \Dzb}}\xspace}
\def\Dp      {\ensuremath{D^+}\xspace}
\def\Dm      {\ensuremath{D^-}\xspace}

\def\DpDm    {\ensuremath{\Dp {\kern -0.16em \Dm}}\xspace}

\def\Bbar    {\kern 0.18em\overline{\kern -0.18em B}{}\xspace}

\def\Bz      {\ensuremath{B^0}\xspace}
\def\Bzb     {\ensuremath{\Bbar^0}\xspace}
\def\BzBzb   {\ensuremath{\Bz {\kern -0.16em \Bzb}}\xspace}
\def\Bu      {\ensuremath{B^+}\xspace}
\def\Bub     {\ensuremath{B^-}\xspace}

\def\BpBm    {\ensuremath{\Bu {\kern -0.16em \Bub}}\xspace}

\def\BorBbar    {\kern 0.18em\optbar{\kern -0.18em B}{}\xspace}
\def\DorDbar    {\kern 0.18em\optbar{\kern -0.18em D}{}\xspace}
\def\KorKbar    {\kern 0.18em\optbar{\kern -0.18em K}{}\xspace}

\mathchardef\Upsilon="7107
\def\Y#1S{\ensuremath{\Upsilon{(#1S)}}\xspace}

\def\FourS {\Y4S}

\mathchardef\Deltares="7101
\mathchardef\Xi="7104
\mathchardef\Lambda="7103
\mathchardef\Sigma="7106
\mathchardef\Omega="710A

\def\Deltabar{\kern 0.25em\overline{\kern -0.25em \Deltares}{}\xspace}
\def\Lbar{\kern 0.2em\overline{\kern -0.2em\Lambda\kern 0.05em}\kern-0.05em{}\xspace}
\def\Sigbar{\kern 0.2em\overline{\kern -0.2em \Sigma}{}\xspace}
\def\Xibar{\kern 0.2em\overline{\kern -0.2em \Xi}{}\xspace}
\def\Obar{\kern 0.2em\overline{\kern -0.2em \Omega}{}\xspace}
\def\Nbar{\kern 0.2em\overline{\kern -0.2em N}{}\xspace}
\def\Xb{\kern 0.2em\overline{\kern -0.2em X}{}\xspace}

\newcommand{\tev}{\ensuremath{\mathrm{\,Te\kern -0.1em V}}\xspace}
\newcommand{\gev}{\ensuremath{\mathrm{\,Ge\kern -0.1em V}}\xspace}
\newcommand{\mev}{\ensuremath{\mathrm{\,Me\kern -0.1em V}}\xspace}
\newcommand{\kev}{\ensuremath{\mathrm{\,ke\kern -0.1em V}}\xspace}
\newcommand{\ev}{\ensuremath{\mathrm{\,e\kern -0.1em V}}\xspace}
\newcommand{\gevc}{\ensuremath{{\mathrm{\,Ge\kern -0.1em V\!/}c}}\xspace}
\newcommand{\mevc}{\ensuremath{{\mathrm{\,Me\kern -0.1em V\!/}c}}\xspace}
\newcommand{\gevcc}{\ensuremath{{\mathrm{\,Ge\kern -0.1em V\!/}c^2}}\xspace}
\newcommand{\mevcc}{\ensuremath{{\mathrm{\,Me\kern -0.1em V\!/}c^2}}\xspace}

\def\invfb   {\ensuremath{\mbox{\,fb}^{-1}}\xspace}

\def\mus  {\ensuremath{\rm \,\mus}\xspace}

\def\ps   {\ensuremath{\rm \,ps}\xspace}
\def\fs   {\ensuremath{\rm \,fs}\xspace}

\def\mus        {\ensuremath{\,\mu{\rm s}}\xspace}    
\def\ps         {\ensuremath{{\rm \,ps}}\xspace}  

\def\to                 {\ensuremath{\rightarrow}\xspace}

\newcommand{\stat}{\ensuremath{\mathrm{(stat)}}\xspace}
\newcommand{\syst}{\ensuremath{\mathrm{(syst)}}\xspace}

\def\pep2{PEP-II}

\newcommand{\dedx}{\ensuremath{\mathrm{d}\hspace{-0.1em}E/\mathrm{d}x}\xspace}
\newcommand{\chisq}{\ensuremath{\chi^2}\xspace}

\def\gsim{{~\raise.15em\hbox{$>$}\kern-.85em
          \lower.35em\hbox{$\sim$}~}\xspace}
\def\lsim{{~\raise.15em\hbox{$<$}\kern-.85em
          \lower.35em\hbox{$\sim$}~}\xspace}

\def\CP                {\ensuremath{C\!P}\xspace}

\xspace

\def\jetset74   {\mbox{\tt Jetset \hspace{-0.5em}7.\hspace{-0.2em}4}\xspace}

\def\Kmpip      {\ensuremath{K^{-}\pi^{+}}\xspace}

\def\KpKm       {\ensuremath{K^{+}K^{-}}\xspace}
\def\Kpi        {\ensuremath{K\pi}\xspace}

\def\pippim     {\ensuremath{\pi^{+}\pi^{-}}\xspace}

\newcommand{\kevcc}{\ensuremath{{\mathrm{\,Ke\kern -0.1em V\!/}c^2}}\xspace}

\def\t          {\ensuremath{t}}
\def\terr       {\ensuremath{\sigma_{\t}}\xspace}

\def\Pchisq      {\ensuremath{P(\chi^2)}}

\def\Dstp        {\ensuremath{D^{*+}}}

\def\ttrue  {\ensuremath{t_{\rm true}}\xspace}
\def\toff   {\ensuremath{t_0}\xspace}

\def\Pbar    {\ensuremath{\kern 0.2em\overline{\kern -0.2em P}{}}\xspace}

\catcode`\@=11\relax
\newskip\dkwidth
\def\dk{%
   \dkwidth=2em plus 0.5 em minus 0.25 em\relax
   {\m@th\mathord{%
   \hbox{%
      \kern 0.3em
      \raise 0.6ex%
      \hbox{%
         \vrule width 0.25pt height 0.5\dkwidth depth0pt}%
      \kern-1.2pt%
      \hbox to 1.1\dkwidth{%
         \rightarrowfill}%
      \kern0.4em}}%
   }%
}%

\def\rightarrowfill{$\m@th\mathord-\mkern-10mu%
  \cleaders\hbox{$\mkern-2mu\mathord-\mkern-2mu$}\hfill
  \mkern-6mu\mathord\rightarrow$}

\catcode`\@=12\relax

\def\yCP        {\ensuremath{y_{C\!P}}\xspace}

\def\Kmpip      {\ensuremath{\Km\pip}\xspace}

\def\tauKpi     {\ensuremath{\tau_{K\pi}}\xspace}
\def\tauhh      {\ensuremath{\tau_{hh}}\xspace}

\def\tauKK      {\ensuremath{\tau_{KK}}\xspace}

\def\DeltayCP   {\ensuremath{\Delta\yCP}\xspace}

\newcommand{\BABARPubYear}    {09}
\newcommand{\BABARPubNumber}  {023}
\newcommand{\SLACPubNumber} {13737}

\def\figurebox#1#2#3{%
    \def\arg{#3}%
    \ifx\arg\empty
    {\hfill\vbox{\hsize#2\hrule\hbox to #2{\vrule\hfill\vbox to
     #1{\hsize#2\vfill}\vrule}\hrule}\hfill}%
    \else
    {\hfill\epsfbox{#3}\hfill}%
    \fi}

\long\def\inst#1{\par\nobreak\kern 4pt\nobreak
    {\it #1}\par\vskip 10pt plus 3pt minus 3pt}

\begin{document}

  \begin{flushleft}
    \rm arXiv:XXXX.XXXX \\
    hep-ex\\
    \today \\[.7in]
  \end{flushleft}

\begin{flushright}
\babar-PUB-\BABARPubYear/\BABARPubNumber\\
SLAC-PUB-\SLACPubNumber
\end{flushright}

\title{
\Large \bf \boldmath
Measurement of $\Dz$-$\Dzb$ Mixing using the Ratio of Lifetimes for the Decays
$\Dz \to \Kmpip$ and $\KpKm$
}

\author{B.~Aubert}
\author{Y.~Karyotakis}
\author{J.~P.~Lees}
\author{V.~Poireau}
\author{E.~Prencipe}
\author{X.~Prudent}
\author{V.~Tisserand}
\affiliation{Laboratoire d'Annecy-le-Vieux de Physique des Particules (LAPP), Universit\'e de Savoie, CNRS/IN2P3,  F-74941 Annecy-Le-Vieux, France}
\author{J.~Garra~Tico}
\author{E.~Grauges}
\affiliation{Universitat de Barcelona, Facultat de Fisica, Departament ECM, E-08028 Barcelona, Spain }
\author{M.~Martinelli$^{ab}$}
\author{A.~Palano$^{ab}$ }
\author{M.~Pappagallo$^{ab}$ }
\affiliation{INFN Sezione di Bari$^{a}$; Dipartimento di Fisica, Universit\`a di Bari$^{b}$, I-70126 Bari, Italy }
% ...
\collaboration{The \babar\ Collaboration}
\noaffiliation

\date{\today}

\begin{abstract}
We measure the rate of \Dz-\Dzb{} mixing with the observable $\yCP=(\tauKpi/\tauKK)-1$, where
\tauKK and \tauKpi are respectively the mean lifetimes of \CP-even $\Dz\to\KpKm$ and
\CP-mixed $\Dz\to\Kmpip$ decays, using
a data sample of $384 \invfb$ collected by the
\babar\ detector at the SLAC PEP-II asymmetric-energy $B$ Factory.
From a sample of $\Dz$ and $\Dzb$ decays where the inital flavor of the
decaying meson is not determined, we obtain
$\yCP = [1.12 \pm 0.26 \stat \pm 0.22 \syst]\%$, which
excludes the no-mixing hypothesis at $3.3 \sigma$, including both
statistical and systematic uncertainties.
This result is in good agreement with a previous \babar\, measurement of $\yCP$
obtained from a sample of $\Dstp\to\Dz\pip$ events,
where the $\Dz$ decays to $\Kmpip$, $\KpKm$, and $\pippim$, which is disjoint
with the untagged $\Dz$ events used here.
Combining the two results taking into account statistical and systematic uncertainties,
where the systematic uncertainties are assumed to be $100\%$ correlated,
we find $\yCP = [1.16 \pm 0.22 \stat \pm 0.18 \syst]\%$,
which excludes the no-mixing hypothesis at $4.1 \sigma$.
\end{abstract}


\vfill

\maketitle

\newpage

Several recent results~\cite{Aubert:2007wf,Staric:2007dt,Abe:2007rd,CDF:2007uc}
show evidence for mixing in the \Dz-\Dzb{}
system consistent with predictions of possible Standard Model
contributions~\cite{Wolfenstein:1985ft,Donoghue:1985hh,Bigi:2000wn,Falk:2001hx,Falk:2004wg}.
These results also constrain many new
physics models~\cite{Burdman:2003rs,Petrov:2006nc,Golowich:2006gq,Golowich:2007ka,Golowich:2009ii}, and
increasingly precise \Dz-\Dzb{} mixing measurements
will provide even stronger constraints.
One manifestation of \Dz-\Dzb{} mixing is differing \Dz\ decay time
distributions for decays to different \CP eigenstates~\cite{Liu:1994ea}.
We present here a measurement of this lifetime difference using
a sample of $\Dz$ and $\Dzb$ decays in which the initial flavor
of the decaying meson is unknown.

Assuming \CP\ conservation in mixing, the two neutral
$D$ mass eigenstates $| D_1 \rangle$ and $| D_2
\rangle$ can be represented as
\begin{equation}
\begin{array}{rcl}
| D_1 \rangle &=& p | \Dz \rangle + q | \Dzb \rangle \\
| D_2 \rangle &=& p | \Dz \rangle - q | \Dzb \rangle \;,
\end{array}
\label{eq:qpdef}
\end{equation}
where $\left|p\right|^2 + \left|q\right|^2 = 1$.
The rate of $\Dz$-$\Dzb$ mixing can be characterized
by the parameters $x \equiv \Delta m/\Gamma$ and $y \equiv \Delta\Gamma/2\Gamma$, where
$\Delta m = m_1 - m_2$ and $\Delta \Gamma = \Gamma_1 - \Gamma_2$
are respectively the differences between the mass and width eigenvalues of the
states in Eq.~(\ref{eq:qpdef}), and
$\Gamma = (\Gamma_1+\Gamma_2)/2$ is the average width.
If either $x$ or $y$ is non-zero, mixing will occur,
altering the decay time distribution of
$\Dz$ and $\Dzb$ mesons decaying into final states of specific
\CP~\cite{Amsler:2008zzb}.

In the limit of small mixing, and no \CP violation in
mixing or in the interference between
mixing and decay (assumptions which are consistent
with current experimental results),
the mean lifetimes of decays
to a \CP eigenstate of a sample
of $\Dz$ ($\tau^{\Dz}_{hh}$)
and $\Dzb$ ($\tau^{\Dzb}_{hh}$),
along with the mean lifetime of decays to
a state of indefinite \CP ($\tauKpi$), can be combined
into the quantity
\begin{equation}
\displaystyle\yCP = \frac{\langle\tauKpi\rangle}{\langle\tauhh\rangle} - 1,
\end{equation}
where $\langle \tauhh \rangle = (\tau^{\Dz}_{hh} + \tau^{\Dzb}_{hh} )/2$.
Noting that the untagged $\Kmpip$~\cite{CC:2008}
final state is a mixture of Cabbibo-favored and doubly Cabbibo-suppressed $\Dz$ and $\Dzb$ decays with a purely
exponential lifetime distribution, along with a very small admixture of mixed $\Dz$
decays, an analogous expression also holds for $\langle \tauKpi \rangle$.
Given the current experimental evidence indicating a small mixing rate,
the lifetime distribution for all $hh$ and $K\pi$ final states is exponential to a good approximation.
If $\yCP$ is zero there is no $\Dz$-$\Dzb$ mixing attributable to a
width difference, although mixing caused by a mass difference may be present.
In the limit of no direct \CP violation, $\yCP = y$.

We measure the \Dz mean lifetime in the \Dz decay modes
$\Kmpip$ and $\KpKm$, where the initial flavor of
the decaying \Dz is not identified (the {\sl untagged} sample).
This sample excludes \Dz mesons which can be reconstructed as
part of $D^{*+} \to \Dz\pip $ decays, as these decays (the {\sl tagged}
sample) are the subject of an earlier \babar\, analysis~\cite{Aubert:2007en}
whose results are combined with those of the current analysis.
To avoid potential bias,
we finalized our data selection criteria,
fitting methodology,
sources of possible systematic uncertainties to be examined,
and method of calculating statistical limits
for the current untagged analysis alone and in
combination with the tagged analysis,
prior to examining the mixing results from the untagged data.
In general, systematic uncertainties related to the reconstruction
of signal events cancel in the lifetime ratio.
However, uncertainties related to the somewhat differing backgrounds
present in the $K^-\pi^+$ and $K^+K^-$ final states lead to larger
systematic uncertainties in the untagged analysis compared to those of
the tagged analysis, which has much higher signal purity .

We use $384\invfb$ of $\epem$ colliding-beam data recorded at, and slightly
below, the $\FourS$ resonance (center-of-mass [CM] energy $\sqrt{s} \sim 10.6\gev$)
with the \babar\ detector~\cite{Aubert:2001tu}
at the SLAC National Accelerator Laboratory \hbox{PEP-II} asymmetric-energy $B$ Factory.
Candidate \Dz signal decays are reconstructed in the final states
$\Kmpip$ and $\KpKm$. The selection of events and reconstruction
of $\Dz$ signal candidates closely follows that of our
previous tagged analysis~\cite{Aubert:2007en}.
We require $\Kp$ and $\pip$ candidates to satisfy
particle identification criteria based on \dedx ionization energy loss
and Cherenkov angle measurements.
We fit oppositely charged pairs of these candidates with appropriate mass hypotheses
to a common vertex to form a \Dz candidate.
The decay time $\t$ of each \Dz candidate with invariant mass within
the range $1.80-1.93~\gevcc$, along with its estimated
uncertainty~$\terr$, is determined from a combined fit to the \Dz production
and decay vertices, with a constraint that the production point
be consistent with the $\epem$ interaction region as determined on
an event-by-event basis.
We retain only candidates with a \chisq-based probability
for the fit $\Pchisq > 0.1\%$, and with $-2 < \t < 4 \ps$
and $\terr < 0.5 \ps$.

We further require the helicity angle $\theta_H$,
defined as the angle between the positively
charged track in the \Dz rest frame and
the \Dz direction in the laboratory frame, to
satisfy $|\cos\theta_H|<0.7$, which aids in the
rejection of purely combinatorial background events.
Contributions from true \Dz mesons
produced in $B$ meson decay are reduced to a negligible amount
by rejecting \Dz candidates with momentum
in the $\epem$ CM frame less than $2.5\gevc$.
For events with multiple candidates sharing one or more tracks,
we retain only the candidate with the highest \Pchisq. The
fraction of events with multiple signal candidates is $\sim 0.05\%$
for the $\KpKm$ final state, and $\sim 0.3\%$ for $\Kmpip$.

\begin{figure}
\begin{center}
\includegraphics[width=0.5\linewidth]{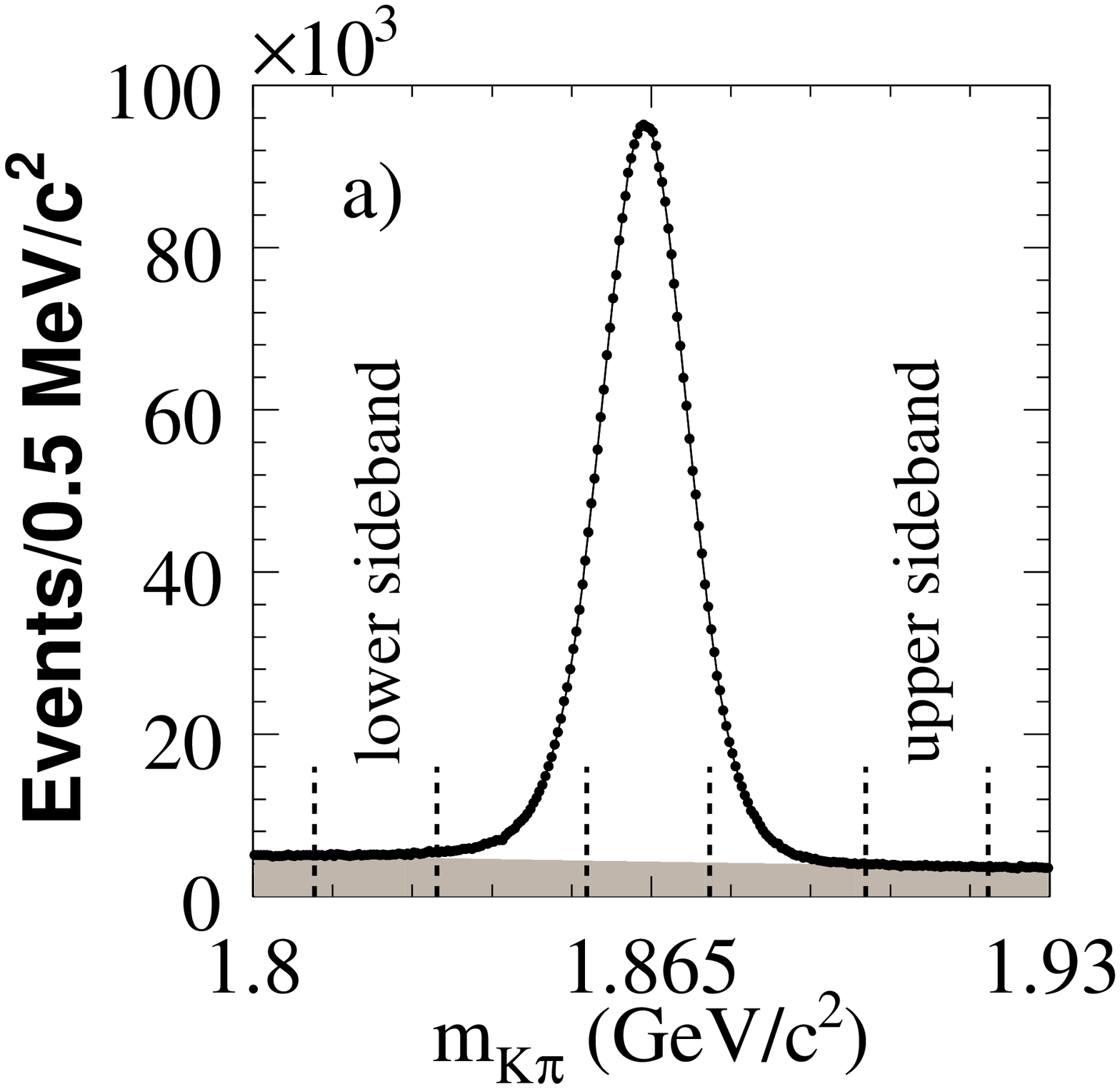}\hfill
\includegraphics[width=0.5\linewidth]{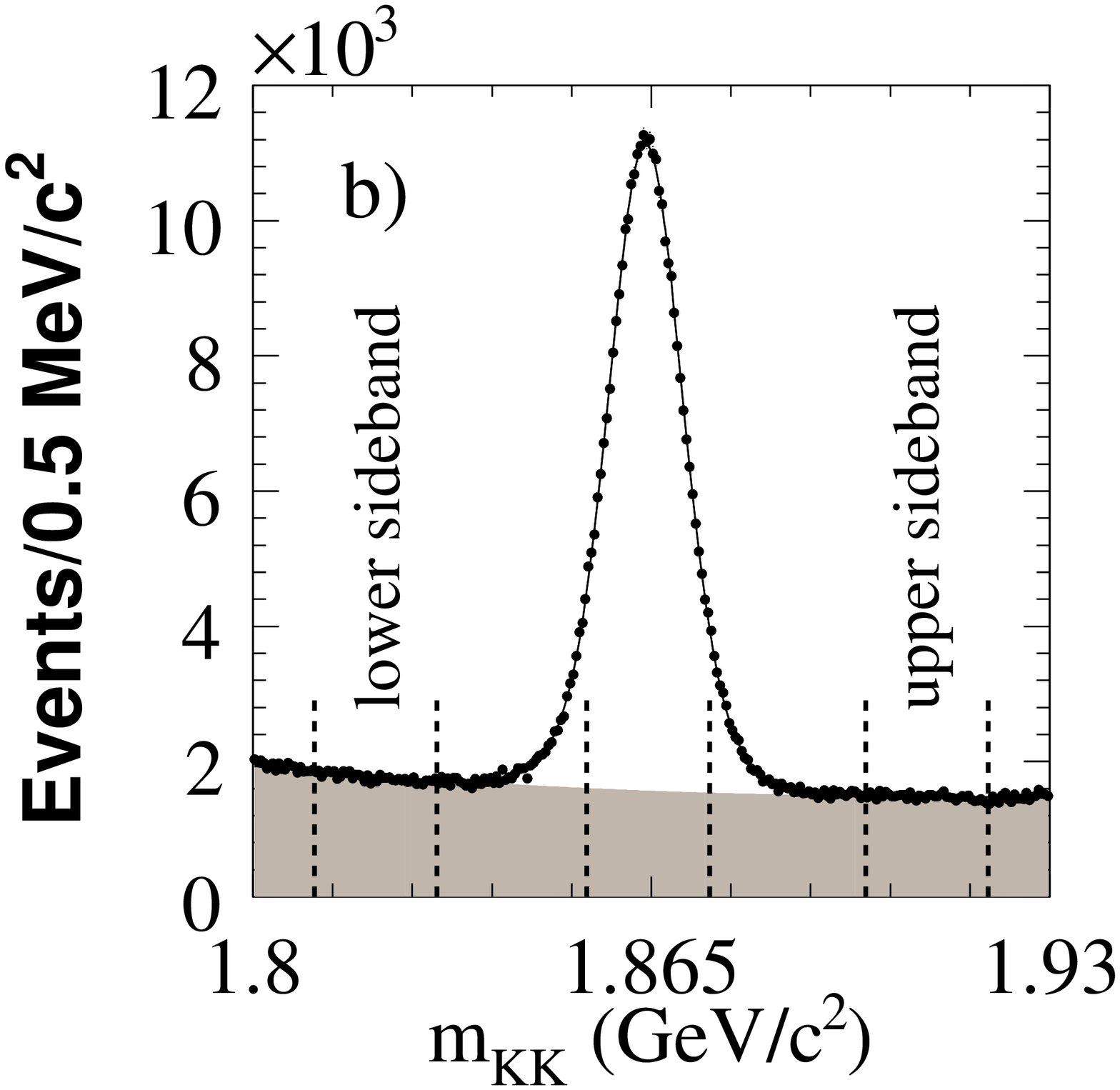}
\caption{(a) $\Dz\to\Kmpip$ and (b) $\Dz\to\KpKm$ invariant mass distribution
with the data (points), total fit (line) and
background contribution (solid) overlaid.
The innermost dashed lines on either side
of the signal peaks delimit the lifetime fit mass region,
with lower and upper mass sidebands shown on either side.}
\label{fig:MassPlot}
\end{center}
\end{figure}

The invariant mass distributions for the final
$\Dz\to\Kmpip$ and $\Dz\to\KpKm$ samples are shown in
Fig.~\ref{fig:MassPlot}.
For the lifetime fits, we use only events
within $\pm 10\mevcc$ of the \Dz
signal peak $1.8545<M_{\Dz}<1.8745\gevcc$ (the {\sl lifetime fit mass region}).
The $\Kmpip$ and $\KpKm$ signal yields within this
region and their purity are given in Table~\ref{tab:nsgnl}.
Events within the mass sideband regions
$1.81<M_{\Dz}<1.83\gevcc$ and $1.90<M_{\Dz}<1.92\gevcc$
are used to determine the combinatorial background
decay time distribution within the lifetime fit mass region.
In addition to purely combinatorial backgrounds,
there are small background contributions from decays
of non-signal charm parents where two of the decay
products are selected as the daughters of a signal
decay and subsequently pass the final event selection.
These misreconstructed charm backgrounds are accounted for
using simulated events.
Their contribution is $\sim 0.7\%$ ($\sim 3.8\%$) of the total number
of background events in the $\Kmpip$ ($\KpKm$) signal
region.

\begin{table}
\centering
\caption{$\Dz\to\Kmpip$ and $\Dz\to\KpKm$ signal
yield and purity in the lifetime fit mass region.}
\begin{tabular}{lcc}
\hline \hline \\
Sample &  Signal Yield (x $10^{3}$) & Purity (\%)
\\ \hline \\
$\Kmpip$    & $2710.2 \pm 3.4$ & 94.2 \\
$\KpKm$     & $263.6  \pm 1.0$ & 80.9 \\
\hline \hline
\end{tabular}
\label{tab:nsgnl}
\end{table}

The mean $\Dz$ lifetime is determined from a fit essentially
identical to the one performed in the previous
tagged analysis~\cite{Aubert:2007en}, using the
reconstructed decay time $t$ and the decay time uncertainty $\terr$
for events within the lifetime fit mass region.
Three categories of events are accounted for in the
lifetime fit: signal decays, combinatorial
background, and misreconstructed charm events.

The decay time distribution of signal events is
described by an exponential convolved with a resolution function
which is taken as the sum of three Gaussian functions with widths
proportional to $\terr$. The functional form of this probability
density function (PDF) for signal events is

\begin{eqnarray}
\label{eq:signaltimeDistribution}
{\cal R}_{X}(t,\terr;\tau_{X}) &=& f_{t3}{\cal D}(t,\terr;S_X s_3,\toff,\tau_{X}) \nonumber \\
      &+& ( 1 - f_{t3})\Big[ f_{t2}{\cal D}(t,\terr;S_X s_2,\toff,\tau_{X})\qquad\\
      &+& ( 1 - f_{t2}) {\cal D}(t,\terr;S_X s_1,\toff,\tau_{X})\Big], \nonumber
\end{eqnarray}

\noindent where $f_{ti}$ (with $i=1 \ldots 3$)
parameterizes the contribution of
each individual resolution function,
$s_i$ is a scaling factor associated with each Gaussian,
$\tau_{X}$ (where $X=\Kpi$, $KK$) is the lifetime parameter determined by the fit,
$\toff$ is an offset to the mean of the resolution function, and where
\begin{equation}
\label{eq:test}
\begin{array}{l}
{\cal D}(t,\terr;s,\toff,\tau) =\\[5pt] 
\quad
C_{\terr}{\displaystyle\int} \exp(-\ttrue/\tau)
                               \exp\left( -\frac{(t-\ttrue+\toff)^2}{2(s\cdot\terr)^2}\right)\, d\ttrue
\end{array}
\end{equation}

\noindent with normalization coefficient $C_{\terr}$.
Up to an overall scale factor in the width,
the resolution function is identical for both final
states.
We account for a small ($\sim 1\%$)
difference in the $\Kmpip$ and $\KpKm$
resolution function width using
an additional fixed scale factor $S_X$.
The value of $S_{KK}$ is determined from
the data, with $S_{\Kpi}$ fixed to 1.0.
Possible biases resulting from this assumption are
included as part of the study of systematic uncertainties.
All other resolution function parameters
are shared among the two modes, and
all parameters are allowed to vary
in a simultaneous extended unbinned maximum
likelihood fit to both final states.

The decay time distribution of the combinatorial background is
described by a sum of two Gaussians and a modified Gaussian with
a power-law tail to account for a small number of events with
large reconstructed lifetimes. The widths of these Gaussians
are not scaled using event-by-event uncertainties.
Events in the lower and upper $\Kmpip$ ($\KpKm$)
mass sidebands are fit separately,
and a weighted average of the results of these fits is used
to parameterize the PDF for $\Kmpip$ ($\KpKm$) combinatorial
events in the lifetime fit mass region.

Misreconstructed charm background events have one or more of the charm decay
products either not reconstructed or reconstructed with the wrong
particle hypothesis.
In the $\Kmpip$ ($\KpKm$) final state, $\sim 60\%$ ($\sim 95\%$)
of these events are from true \Dz decays, with the balance coming
from charged $D$ and charm baryon decays.
The charm background is long-lived and is described using an exponential
convolved with a resolution function consisting of two Gaussians with a
shared mean and widths that depend on \terr.
Because the number of these events in the $\Kmpip$ ($\KpKm$) sample is small
relative to the total background, an effective
lifetime distribution taken from simulated events and summed over all
$\Kmpip$ ($\KpKm$) charm backgrounds is used in the $\Kmpip$ ($\KpKm$)
lifetime fit.

Since the lifetime fit PDFs depend on the event-by-event
decay time uncertainty, PDFs describing the distribution of decay
time uncertainties for each of the event classes are required
to avoid bias in the likelihood estimator used in the data fit~\cite{Punzi:2004wh}.
We extract these distributions directly from the data.
For combinatorial events, the distribution of decay time uncertainties
is taken from a weighted average of the distributions extracted from the
lower and upper mass sidebands.
The decay time uncertainty distribution for signal events is obtained by
subtracting the combinatorial background uncertainty distribution from the
uncertainty distribution of all ({\sl i.e.}, background plus signal) candidates
present in the lifetime fit mass region.
The signal distribution is also used for the relatively small number
of misreconstructed charm background events.

\begin{figure}
\begin{center}
\includegraphics[width=0.95\linewidth, clip=]{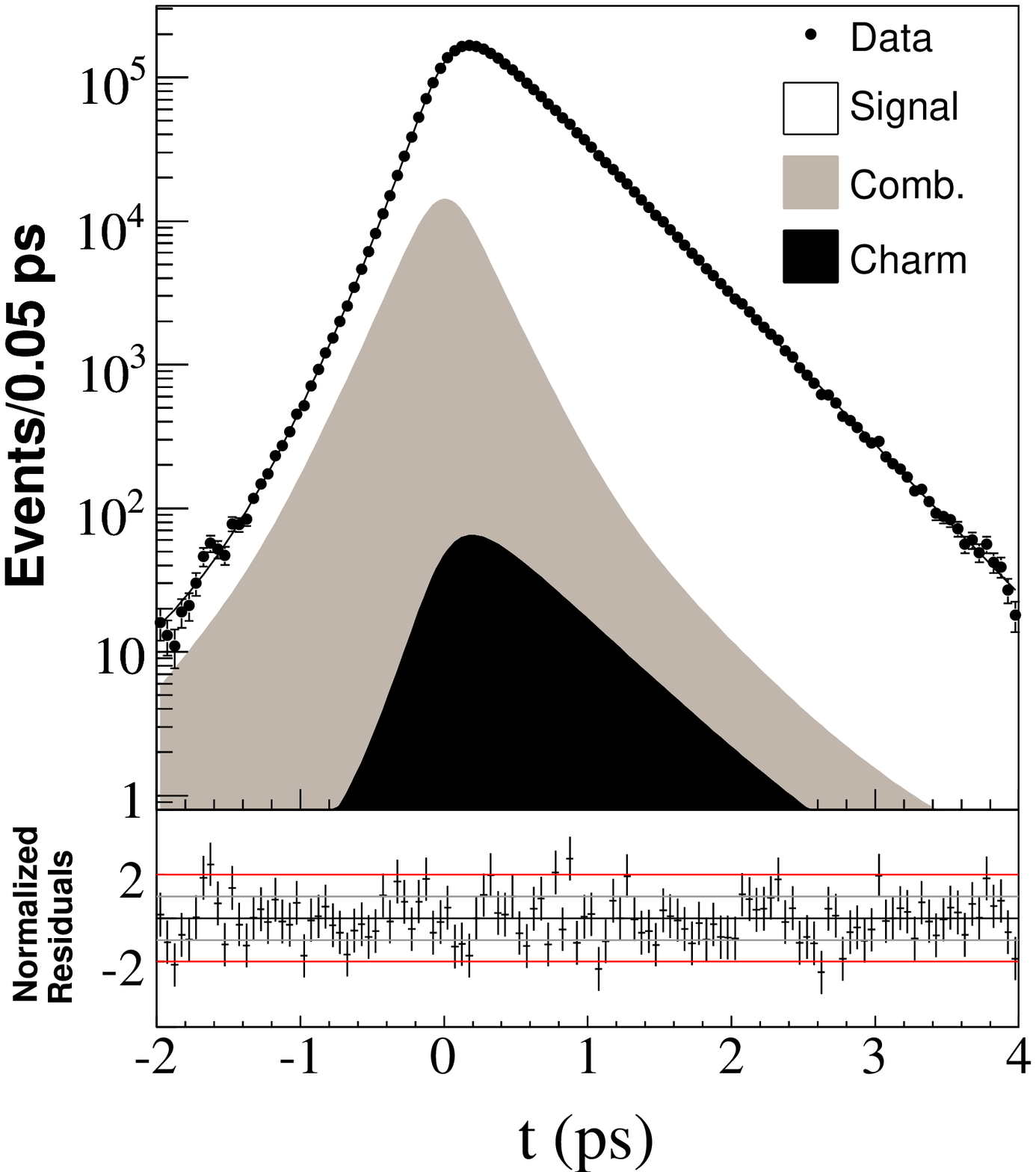}\hfill
\caption{$\Dz\to\Kmpip$ decay time distribution with the data (points), total lifetime fit (line),
combinatorial background (gray) and charm background (black) contributions overlaid.}
\label{fig:KpiDecayTime}
\end{center}
\end{figure}

\begin{figure}
\begin{center}
\includegraphics[width=0.95\linewidth, clip=]{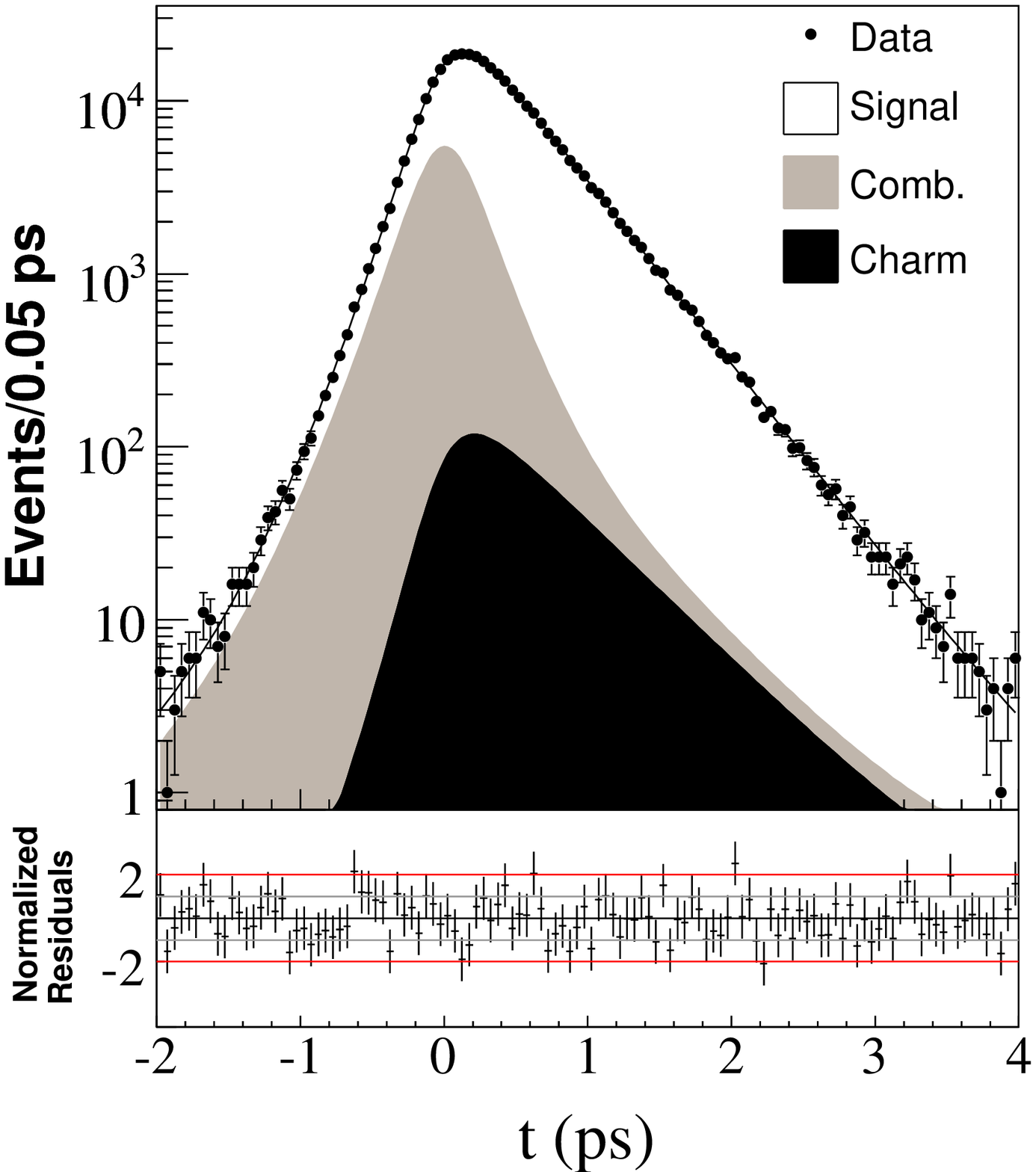}\hfill
\caption{$\Dz\to\KpKm$ decay time distribution with the data (points), total lifetime fit (line),
combinatorial background (gray) and charm background (black) contributions overlaid.}
\label{fig:KKDecayTime}
\end{center}
\end{figure}

The results of the lifetime fits are shown in
Figs.~\ref{fig:KpiDecayTime} and~\ref{fig:KKDecayTime},
along with a plot of the point-by-point residuals for each fit
normalized by the statistical uncertainty associated with a data point.
We find the $\Dz\to\Kmpip$ mean lifetime
$\tau_{\Kpi} = 410.39 \pm 0.38 \stat \fs$ and the $\Dz\to\KpKm$ mean lifetime
$\tau_{KK} = 405.85 \pm 1.00 \stat \fs$, yielding $\yCP = [1.12 \pm 0.26 \stat ]\%$.
The statistical significance of this mixing result without taking into account
systematic uncertainties is $4.3 \sigma$. This untagged result is in good
agreement with our previous tagged analysis~\cite{Aubert:2007en}. When the
two results are combined, we find $\yCP = [1.16 \pm 0.22 \stat]\%$,
a result with a statistical significance of $5.3 \sigma$, excluding
any systematic uncertainties.

Numerous cross-checks have been performed to assure the unbiased nature
of the fit model and to validate the assumptions used in its construction.
We have performed fits to datasets composed of fully simulated signal
and background events in the proportions seen in the actual data, and
find no bias in the measurement of individual $\tau_{\Kpi}$ and $\tau_{KK}$
lifetimes for simulated signal events generated at $411.6 \fs$ (very near
the nominal $\Dz \to \Kmpip$ lifetime value~\cite{Amsler:2008zzb}), or for a lifetime value
$\sim 10\%$ greater than this for $\Dz \to \KpKm$ . We additionally find no significant variations
in the reconstruction efficiency for signal decays as a function of the true decay time.

Many of the systematic uncertainties associated with the
individual lifetime measurements cancel to a great extent
in the ratio of lifetimes. We consider as possible sources
of systematic uncertainty: variations in the
signal and background fit models, changes to the event
selection, and detector effects that might introduce
biases in the lifetime measurements.

We test the assumption of a shared signal resolution model by separately
fitting each mode using completely independent resolution functions, and
assign as a systematic uncertainty the magnitude of the change $|\DeltayCP|$ in $\yCP$ relative
to the result of the nominal fit. We additionally perform the nominal fit
using a double Gaussian signal resolution model, and similarly assign a
systematic uncertainty.
The total uncertainty associated with the choice of
signal resolution model is 0.016\%.

To estimate possible biases correlated with the extent and position of
the lifetime fit mass region, the size of the mass window is varied by
$\pm 2$ and $\pm 5 \mevcc$ without changing the mass region center, and the
center is shifted by $\pm 0.5 \mevcc$ while retaining the nominal
$20 \mevcc$ width.
The total systematic uncertainty obtained from variations in the lifetime fit
mass window is 0.110\%.

The modeling of the misreconstructed charm background is taken from
simulated events, and we vary the expected contribution from these
events by $\pm 15\% (\pm 5\%)$ for the $\Kmpip$ ($\KpKm$) final state.
These bounds are conservatively assigned based on the results of other
\babar\, charm analyses in which the background modes here are fully
reconstructed, and in which data and simulated event yields are found to agree
within a few percent. We additionally vary the effective lifetime used
in the charm background lifetime fit PDFs by the same percentages, which corresponds to
$> \pm 2 \sigma$ in the statistical uncertainty given the number of
simulated events used. The largest $|\DeltayCP|$ value within each of these two
classes of variations is assigned as a systematic uncertainty, 0.0585\% for the
normalization variations and 0.0624\% for the effective lifetime variations, which
are then added in quadrature.

We account for a possible bias associated with obtaining the
combinatorial lifetime PDF in the lifetime fit mass region from
data in the lower and upper mass sidebands by fluctuating the PDF
parameters taking into account the correlations and statistical
errors resulting from the sideband fits.
We construct 100 PDF variations for each of the lower and upper sidebands for each
of the final states, and then perform the nominal lifetime fit using
each variation. We separately compute the RMS of the 100 $\DeltayCP$
values associated with each of the four sets of variations, and
assign the largest RMS, 0.115\%, as a systematic uncertainty.

We evaluate systematic uncertainties associated with the selection of
the final dataset by individually varying the selection criteria.
We change the maximum allowed decay time uncertainty by $\pm 0.1 \ps$,
and assign the largest $|\DeltayCP|$ value, 0.069\%, as a systematic uncertainty.
We vary the way in which signal candidates that share
tracks with other signal candidates are selected by removing
all overlapping candidates, and separately by also retaining all such
candidates, and again take the larger of the resulting two
$|\DeltayCP|$ values, 0.017\%, as a systematic uncertainty.

We account for possible detector effects which might bias the
lifetime ratio by using several different detector configurations
to re-reconstruct simulated event samples with statistics greater
than the actual data for each configuration.
These configurations include vertex detector misalignments, along with boost
and beamspot variations, whose extent is based on residual uncertainties
in studies of mu-pair and cosmic events.
The misalignment configurations introduce changes of up to $4\fs$ in both $KK$ and $K\pi$ lifetimes,
as well as changes in the offset parameter $t_o$  of up to $5\fs$.
Since the same simulated event sample is reconstructed for each set of
detector configuration, the variations are dominated by systematic effects.
The total systematic uncertainty arising from this source is 0.093\%.

\begin{table}
\centering
\caption{Systematic uncertainties.}
\begin{tabular}{lc}
\hline \hline \\
Uncertainty Source            & $|\Delta\yCP|$ (\%)
\\ \hline \\
Signal resolution model       &     0.016   \\
Mass window                   &     0.110   \\
Misreconstructed Charm model  &     0.086   \\
Combinatorial PDF             &     0.115   \\
$\terr$ selection             &     0.069   \\
Overlap candidate selection   &     0.017   \\
Detector effects              &     0.093   \\
\hline \\
Total             &     0.216   \\
\hline \hline
\end{tabular}
\label{tab:syserr}
\end{table}

Table~\ref{tab:syserr} shows the contribution from each source of
systematic uncertainty given above. The total is calculated as the sum in
quadrature of each of the individual items. In addition to the
contributions quantified in the table, we also look for
possible biases by fitting the data separated in: several
different data-taking periods; several different azimuthal and polar
angle bins in the laboratory frame for the $\Dz$ candidate; several
bins of the opening angle in the laboratory frame between the two
$\Dz$ daughters; several bins of the $\Dz$ helicity angle; and
several bins of the $\Dz$ momentum in the CM frame. We observed
no significant biases in any of these cases.

In our previously published tagged analysis~\cite{Aubert:2007en},
we combined the tagged result with the result
of an untagged \babar\, analysis done using a much
smaller dataset~\cite{Aubert:2003pz}, and this previous
untagged result is superseded by the result here, which is
$\yCP({\hbox{untagged}}) = [1.12 \pm 0.26 \stat \pm 0.22 \syst]\%$, which
excludes the no-mixing hypothesis at $3.3 \sigma$, including both
statistical and systematic uncertainties.
Our previous tagged result~\cite{Aubert:2007en} is
$\yCP({\hbox{tagged}}) = [1.24 \pm 0.39 \stat \pm 0.13 \syst]\%$.
These results contain no events in common, and are thus statistically
uncorrelated by construction. However, the degree of correlation in the
systematic uncertainties is substantial, and we conservatively assume
a $100\%$ correlation in the systematics shared between the two analyses.
Combining the tagged and untagged results
taking into account both statistical and systematic
uncertainties~\cite{Lyons:1988rp}, we find
$\yCP({\hbox{correlated}}) = [1.16 \pm 0.22 \stat \pm 0.18 \syst]\%$.
Summing statistical and systematic uncertainties in quadrature, the
significance of this measurement is $4.1 \sigma$.

We are grateful for the
extraordinary contributions of our \hbox{PEP-II} colleagues in
achieving the excellent luminosity and machine conditions
that have made this work possible.
The success of this project also relies critically on the
expertise and dedication of the computing organizations that
support \babar.
The collaborating institutions wish to thank
SLAC for its support and the kind hospitality extended to them.
This work is supported by the
US Department of Energy
and National Science Foundation, the
Natural Sciences and Engineering Research Council (Canada),
the Commissariat \`a l'Energie Atomique and
Institut National de Physique Nucl\'eaire et de Physique des Particules
(France), the
Bundesministerium f\"ur Bildung und Forschung and
Deutsche Forschungsgemeinschaft
(Germany), the
Istituto Nazionale di Fisica Nucleare (Italy),
the Foundation for Fundamental Research on Matter (The Netherlands),
the Research Council of Norway, the
Ministry of Education and Science of the Russian Federation,
Ministerio de Educaci\'on y Ciencia (Spain), and the
Science and Technology Facilities Council (United Kingdom).
Individuals have received support from
the Marie-Curie IEF program (European Union) and
the A. P. Sloan Foundation.

\end{document}